\documentclass[%
 reprint,
 amsmath,amssymb,
 aps,
]{revtex4-1}
\usepackage{amsmath}
\usepackage{graphicx}
\usepackage{dcolumn}
\usepackage{bm}
\usepackage{hyperref}

\begin{document}

\preprint{APS/123-QED}

\title{Quasi-phase-matching of high-order-harmonic generation using multimode polarization beating}

\author{Lewis Z. Liu}
 \email{L.Liu1@physics.ox.ac.uk}
\author{Kevin O'Keeffe}
\author{Simon M. Hooker}
\affiliation{%
 Clarendon Laboratory, University of Oxford Physics Department,
 Parks Road, Oxford OX1 3PU, United Kingdom
}%

\date{\today}

\begin{abstract}
The generalization of polarization beating quasi-phase matching
(PBQPM) and of multi-mode quasi-phase matching (MMQPM) for the
generation of high-order harmonics is explored, and a novel method
for achieving polarization beating is proposed. If two (and in
principle more) modes of a waveguide are excited, modulation of the
intensity, phase, and/or polarization of the guided radiation will
be achieved; by appropriately matching the period of this modulation
to the coherence length, quasi-phase-matching of high harmonic
radiation generated by the guided wave can occur.  We show that it
is possible to achieve efficiencies with multi-mode quasi-phase
matching greater than the ideal square wave modulation. We present a
Fourier treatment of QPM and use this to show that phase modulation,
rather than amplitude modulation, plays the dominant role in the
case of MMQPM. The experimental parameters and optimal conditions
for this scheme are explored.

Please note that this is an arXiv version of the original APS paper.  Please cite original paper L. Z. Liu, K O'Keeffe, and S. M. Hooker, Phys.  Rev. A 87, 023810 (2013).  APS link here: \url{http://pra.aps.org/abstract/PRA/v87/i2/e023810}
\begin{description}

\item[PACS numbers]
42.55.Vc 42.81.Gs 42.65.Ky

\end{description}
\end{abstract}

\maketitle


\section{Introduction}

High harmonic generation (HHG) is a nonlinear process that enables
the production of tunable, coherent soft x-rays with applications in
time-resolved science \cite{Uiberacker, Schultze,Cavalieri},
ultrafast holography \cite{RaHolography}, or diffractive imaging
\cite{Sandberg}. An important feature of HHG is that it is simple to
achieve: focusing driving laser radiation to an intensity of order
$10^{14}$ W/cm$^2$ in a gaseous target yields coherent radiation
with frequencies corresponding to the odd harmonics of the driving
radiation $\omega$. A semi-classical theory of this phenomena has
been developed by Corkum \cite{Corkum} and a quantum treatment has
been given by Lewenstein et al.\ \cite{Lewenstien}.

However, without additional techniques, HHG is highly inefficient
\--- with a typical conversion ratio of $10^{-6}$ for generating
photons of energy in the 100 eV range, and $10^{-15}$ for generating
1 keV photons.  This low efficiency is partially due to fact that
the driving field and the harmonic field have different phase
 velocities. As a consequence a phase difference develops between the driving field
and harmonics generated at each point in the generating medium; this
in turn causes the intensity of the generated harmonics to oscillate
with distance between zero and some maximum value along the
direction of propagation, $z$.  The phase velocity difference is
characterized by the wave vector mismatch, $\Delta k$, which arises
from neutral gas, plasma, and waveguide dispersion; it is given by
$\Delta k = k(q \omega) - q k(\omega)$ where $q$ is the harmonic and
$k(\omega)$ is the propagation constant for radiation of angular
frequency $\omega$. The distance it takes for the two fields to slip
in phase by $\pi$ is the coherence length $L_c = \pi/\Delta k$.

One way of avoiding the phase-mismatch problem is to balance the
dispersion so that $\Delta k =0$, a situation we will describe as
\lq \lq true phase-matching'' in order to distinguish it from the
quasi-phase-matching discussed below. With true phase-matching --
assuming absorption can be neglected -- the intensity of the
harmonics grows quadratically with the propagation distance $z$
\cite{Murnane1999, Murnane1998, MurnaneScience2012}.  With long wavelength drivers for phase matching, conversion efficiency can be achieved up to $10^{-3}$ for the VUV region and $10^{-6}$ in the x-ray region \cite{MurnaneScience2012}.  However, without long wavelength drivers, true
phase-matching can only be achieved for relatively low-order
harmonics --- corresponding to low photon energies --- since at the
higher driving intensities required to generate high-order harmonics
the dispersion becomes dominated by the free electrons and cannot be
balanced by the other terms \cite{Murnane1999}. For higher-order
harmonics, so-called quasi-phase-matching (QPM) may be employed in
which the harmonic generation is suppressed in the out of phase
zones, enabling monotonic growth of harmonic intensity as a function
of $z$. Techniques for QPM include counter-propagating pulses
\cite{Robinson2010, Dromey, Peatross, Lytle, Zhang}, multi-mode
beating \cite{Zepf2007, RobinsonThesis, DromneyMMQPM, WalterMMQPM},
and modulated waveguides \cite{ModWaveguide}.

Multi-mode QPM (MMQPM) relies on coupling in two or more
waveguide modes \cite{Zepf2007, RobinsonThesis, DromneyMMQPM}.  If
the two modes travel at different phase velocities, then the
intensity will beat along the propagation length, thereby modulating
harmonic generation resulting in QPM . In this paper, we investigate
the effect on HHG of the modulation in both the intensity \emph{and}
phase of the beating driving radiation.  We show that under certain
conditions, phase modulation due to mode beating enables harmonics
to be generated with greater efficiency than ideal square-wave QPM
modulation. Moreover we show that MMQPM is dominated by modulation
of the phase, rather than the intensity, of the driving radiation
--- an effect which was not considered in our earlier analysis
\cite{Zepf2007, RobinsonThesis, DromneyMMQPM}.

Recently we have proposed a new class of QPM based on modulation of
the polarization state of the driving radiation within a hollow core
waveguide \cite{LiuPRAPBQPM, LiuOptLettORQPM, PatentPBQPM, PatentORQPM}. Here we discuss one
example of polarization-control QPM: polarization beating QPM
(PBQPM) \cite{PatentPBQPM, LiuPRAPBQPM}. In this approach, a linear
birefringent system modulates the polarization of the driving pulse
between linear and elliptical. Because harmonic generation is
suppressed for elliptically polarized light, QPM can be can be
achieved if the period of the polarization beating is suitably
matched to the coherence length. This paper describes a
generalization of MMQPM and PBQPM which combines these two schemes:
multi-mode polarization beating quasi-phase matching (MMPBQPM),
which utilizes beating between two waveguide modes to modulate the
intensity, phase, and/or polarization of the guided radiation. These
modulations can lead to QPM if the coherence length $L_c$ of the
harmonic generation is appropriately matched to the beat length,
$L_b$, which is the distance it takes for the two modes to develop a
phase difference of $\pi$.  In addition to controlling the relative
input polarizations of the two modes, the relative polarization
angle between the two modes can also be controlled. This increased
parameter space affords greater opportunities for QPM.

In addition we further analyze MMPBQPM using a Jacobi-Anger and
Fourier decomposition which affords additional insight into the
processes leading to QPM. Similar Fourier techniques used to analyze
quasi-phase matching for HHG can be found in \cite{Bahabad,
Shkolnikov}.

The paper is organized as follows: in Sec. \ref{sec:analysis}, the equations for MMPBQPM are derived
for two modes, Sec \ref{sec:fourier} presents the the Jacobi-Anger and Fourier analytical analysis, and Sec. \ref{sec:simulations} discusses numerical simulation
results.

\section{Derivation of the Envelope Equation \label{sec:analysis}}
\subsection{Mode Propagation Equations \label{sec:modes}}
In this section, we develop the set of general mode propagation
equations for two linearly polarized modes with azimuthal symmetry.
If we assume that two modes are excited, then within the waveguide
the electric field may written in cylindrical coordinates as,
\begin{eqnarray}\vec{\mathfrak{E}}({r}, z) &=& E_0 \left\{c_1 \vec{E_1}({r}) e^{i[(\beta_1 + i \alpha_1)z - \omega t]} \right. \nonumber \\
 &&+ \left. c_2 \vec{E_2}({r}) e^{i[(\beta_2 + i \alpha_2)z - \omega t]} \right\}\end{eqnarray}
where $r$ is the radial coordinate from the propagation axes,
$\vec{E_1}$ and $\vec{E_2}$ are the normalized transverse electric
fields of the driving mode $m=1$ and modifying mode $m=2$
respectively, $E_0$ is the electric field amplitude constant,
$\beta_m$ and $\alpha_m$ are the propagation constant and damping
rate of mode $m$, and $\omega$ is the angular frequency of the
driving radiation.

Here, the polarizations of the modes $m=1$ and $m=2$ are respectively taken to be parallel to, and at an angle $\Omega$, to the x-axis:
\begin{equation} \vec{E_1}({r}) = \left( \begin{array}{ll} M_1({r}) \\ 0 \end{array}
\right) \end{equation}
\begin{equation} \vec{E_2}({r}) = \left( \begin{array}{ll} M_2 ({r}) \cos(\Omega) \\ M_2 ({r}) \sin(\Omega) \end{array}
\right) \end{equation} where $M_m({r})$ is the normalized transverse
electric field profile for the $m$th mode.

At time $t' = t + \Delta t$, where $\Delta t = - \beta_1 / \omega
z$, the electric field components are given by,
\begin{eqnarray} \Re\left[
\left ( \begin{array}{ll} \mathfrak{E}_x \\
\mathfrak{E}_y
\end{array}
\right) \right] &=& E_0\left( \begin{array}{ll}  c_1 M_1({r})  \cos(\omega t') \\
0
\end{array} \right) \\
&&+E_0\left( \begin{array}{ll}   c_2 M_2({r})  \cos(\Omega)
\cos(\omega t' + \Delta \beta z) \\
 c_2 M_2({r}) \sin(\Omega) \cos(\omega t' +\Delta \beta
z) \end{array} \right).  \nonumber
\end{eqnarray} where $\Delta \beta = \beta_1 - \beta_2$, and the beat
length $L_b =\pi/(\Delta \beta)$.

From this, the values of $t'$ corresponding to the maximum and
minimum electric field amplitudes can be found, from which the
ellipticity is given by,
\begin{equation} \varepsilon({r}, z) = \sqrt{\frac{\Re\{\mathfrak{E}[t_{max}'({r}, z); r, z]\}^2}{\Re\{\mathfrak{E}[t_{min}'({r}, z); {r},
z] \}^2}} \end{equation} where $\Re[\mathfrak{E}]^2 =
\Re[\mathfrak{E}_x]^2 + \Re[\mathfrak{E}_y]^2$, and the angle of the
major axis is given by,
\begin{equation} \Theta({r}, z) = \tan^{-1} \left(\frac{\Re[\mathfrak{E}_y][t_{max}'({r}, z); {r}, z]}{\Re[\mathfrak{E}_x][t_{max}'({r}, z); {r},
z]} \right) .\end{equation}

It is useful to note that the relative relative intensity of the
driving wave at any given point is given as:
\begin{eqnarray} I({r}, z) &=& \mathfrak{E}_x({r}, z) \mathfrak{E}_x^*({r}, z) + \mathfrak{E}_y({r},z)\mathfrak{E}_y^*({r},z) \\
&\approx& c_1^2M_1 ^2 + c_2^2M_2^2 \nonumber \\
&&+ 2 c_1M_1  c_2M_2 \cos(\Omega) \cos(\Delta \beta z)
\end{eqnarray} assuming that the damping terms are
small.  For the remaining of the paper, we will focus on $r=0$ and
will define $M_1 = M_1(0)$, $M_2 = M_2(0)$ to simplify notation.
Moreover, to simplify arguments, we will stipulate the following
normalization condition: $c_1^2 M_1^2 + c_2^2 M_2^2 = 1$ where we
have taken $c_m M_m$ to be real, which can always be achieved by a
suitable shift of the $z$ or $t$ coordinates.

\subsection{Derivation of the growth equation}
If we write the electric field of the $q$th harmonic as,
\begin{eqnarray}
\vec{E} &=& \frac{1}{2} \vec{\xi}^{(q\omega)} e^{ i \left[k(q\omega)
z -  q\omega t\right]} + \mathrm{c.c}
\end{eqnarray}
then, within the slowly-varying envelope approximation, the equation
for the growth of the amplitude of the $q$th harmonic becomes,
\begin{eqnarray}
2 i k(q\omega) e^{i k(q\omega) z } \frac{d \vec{\xi}^{(q\omega)}}{d
z} = - \mu_0 (q \omega)^2 \vec{P}_\mathrm{NL}^{(q \omega)}
\end{eqnarray}
where $\vec{P}_\mathrm{NL}^{(q \omega)}$ is the component of the
non-linear polarization oscillating with angular frequency
$q\omega$. Now, $\vec{P}_\mathrm{NL}^{(q \omega)} =
\vec{F}'(I,\epsilon)e^{i q \phi_p(z)}$, where $\vec{F}'(I,\epsilon)$
gives the dependence of the nonlinear response on the intensity and
ellipticity of the driving field, and $\phi_p(z)$ is the phase of
the driving field of the $p$th polarization component. We may write
$\phi_x(z) = k_1(\omega)z + \psi_x(z)$ and $\phi_y(z) = k_2(\omega)z
+ \psi_y(z)$, where $\psi_p(z)$ is the additional phase arising from
interference of the waveguide modes and $k_m(\omega) =
\beta_m(\omega)$ is the waveguide propagation constant for the
driving pulse. Henceforth all equations will refer to the $q$th
harmonic, and so in order to avoid clutter we will drop the $(qw)$
superscripts.  The growth equation for the amplitude of the harmonic
for the x- and y- components may then be written in the form,
\begin{eqnarray}
\frac{d{\xi_x}}{d z}&=&  {F_x}(I,\varepsilon)e^{-i \Delta k_1 z}
e^{i \psi_x(z)} =
{\mathfrak{P}_x}\left|\vec{F}\right| e^{+i \Psi_x(z)} \\
\frac{d{\xi}_y}{d z} &=&  {F_y}(I,\varepsilon)e^{-i \Delta k_2 z}
e^{i \psi_y(z)} =
{\mathfrak{P}_y}\left|\vec{F}\right| e^{+i \Psi_y(z)}
\label{eqn: diffeq1}
\end{eqnarray}
where $\Psi_x$ and $\Psi_y$ are the total phase for the x- and y-
components respectively; $\vec{\mathfrak{P}}$ is a projection term
that relates the nonlinear polarization to x- and y- components of
the envelope function, as developed below; $\Delta k_m = k(q\omega)
- q k_m(\omega)$; and $\vec{F}(I,\epsilon) = i \mu_0 (q\omega)^2
\vec{F'}(I,\varepsilon)/2 k(q\omega)$. We note that, as discussed
below, $\vec{F}(I,\varepsilon)$ is in general complex since the
phase of the nonlinear polarization depends on the trajectory of the
ionized electron, and therefore on both the intensity and
ellipticity of the driving field.  In the equation above, we have
factored these phase terms into $\Psi_x$ and $\Psi_y$.

Considering first the $x$-polarization, the driving field may be
written as,
\begin{eqnarray}
E_x(\vec{r},z,t) &=& \left[c_1 M_1  + c_2 M_2 \cos \Omega e^{-i \Delta \beta z}\right]e^{i (\beta_1 z - \omega t)}\nonumber\\
&=&u_x(z)e^{i (\beta_1 z - \omega t)}
\end{eqnarray}
where $\Delta \beta = \beta_1 - \beta_2$. Hence we find,
\begin{eqnarray}
\psi_x(z) &=& \arg[u_x(z)] \nonumber \\
&=& \tan^{-1} \left[\frac{-c_2 M_2 \cos \Omega \sin(\Delta \beta
z)}{c_1 M_1 + c_2 M_2 \cos \Omega \cos (\Delta \beta
z)}\right] \nonumber \\
&\approx& -\frac{c_2M_2}{c_1M_1} \cos \Omega \sin(\Delta \beta z)
\label{Eqn:Driver_phase_x}
\end{eqnarray} where in the last step, the approximation is valid if $| c_1 | \gg |c_2|$.
Similar considerations show that $\psi_y = 0$ and $\phi_y(z) =
\beta_2$.

The strength of the nonlinear polarization, $|\vec{F}(I,\epsilon)|$,
depends on the intensity and polarization of the driving laser
field. Evaluation of $|\vec{F}(I,\epsilon)|$ requires a model of the
interaction of the driving field with the atom, as, for example,
developed by Lewenstein et al.\ \cite{Lewenstien}. However, for our
purposes it is sufficient to assume that the amplitude of the
nonlinear polarization, $\left | \vec{F} \right|$ can be written in
the form:
\begin{eqnarray}\left|\vec{F}[I(z), \varepsilon(z)] \right| = A G[I(z)] H[\varepsilon
(z)]\end{eqnarray} where $A \propto \mu_0(q \omega)^2/2k(q \omega)$
is a constant, $G[I(z)]$ is the intensity-dependent term ranging
from $[0,1]$, and $H[\varepsilon (z)]$ is the ellipticity-dependent
term ranging from $[0,1]$.

For the purposes of illustrating the operation of MMPBQPM it is
sufficient to assume that the intensity-dependent term takes the
form of a power law $G(I) \approx I^{\eta/2}$.  We will assume $\eta
\approx 6$, in accordance with earlier work \cite{DromneyMMQPM,
Antoine,  Antoine2}; but note that the the broad conclusions of the
present paper do not depend strongly on the value of $\eta$.

It is also well known that the single-atom efficiency of HHG depends
sensitively on the polarization of the driving laser field
\cite{Budil1993} which arises from the fact that the ionized
electron must return to the parent ion in order to emit a harmonic
photon.  Following the argument given in \cite{LiuPRAPBQPM}, for a
given driving intensity the number of harmonic photons generated as a function of ellipticity maybe approximated by:
\begin{equation}h(\varepsilon) \approx \left( \frac{1 + \varepsilon^2}{1-
\varepsilon^2}\right) ^{\mu} \label{eqn:budil} \end{equation}
where $H(\varepsilon) = \sqrt{h(\varepsilon)}$.

It is predicted that $\mu = q-1$ within the perturbative regime, as
verified \cite{Budil1993} by Budil et al.\  for harmonics $q = 11$
to $19$,  and by Dietrich et al.\   for harmonics up to $q\approx
31$ \cite{DietrichPolarization}. Schulze et al.\  found that for
higher-order harmonics the sensitivity of harmonic generation to the
ellipticity of the driving radiation is lower than predicted by Eqn
\eqref{eqn:budil} with $\mu = q-1$ \cite{SchulzePolarization},
although in this non-perturbative regime the efficiency of harmonic
generation still decreases with $\varepsilon$. Further measurements
of the dependence of harmonic generation on $\varepsilon$ have been
provided by Sola et al.\ \cite{SolaPolarization}.   It is recognized
that Eqn \eqref{eqn:budil} is an approximation, but it will serve
our purpose of demonstrating the operation of MMPBQPM.

The offset angle and ellipticity of the harmonics generated by
elliptically-polarized radiation have been shown to depend on the
ellipticity and intensity of the driving radiation, and on the
harmonic order \cite{Soviet2, Antoine,  Antoine2, Strelkov,
SchulzePolarization}. Propagation effects can also play an important
role. Since the amplitude with which harmonics are generated
decreases strongly with increasing ellipticity, we are most
interested in the ellipticity of the harmonics generated for small
$\varepsilon$. It has been shown that for higher-order harmonics,
and/or high driving intensities, both the ellipticity and change in
ellipse orientation of the harmonics generated by radiation with
$\varepsilon \approx 0$ are close to zero \cite{Antoine}. We will
therefore make the simplification that the generated harmonics are
linearly polarized along the major axis of the driving radiation and
that we resolve separately the harmonics polarized along the fast
and slow axes of the waveguide.  Thus, the projection term may be
written as:
\begin{equation}
\vec{\mathfrak{P}}(z) = \left( \begin{array}{ll} \cos[\Theta(z)] \\
\sin[\Theta(z)] \end{array} \right), \end{equation} and by following
the arguments of \cite{LiuPRAPBQPM}, the coherence lengths for
harmonics polarized parallel to the x- and y- axes are different,
and hence for a given $L_b$ it is only possible to quasi-phase-match
one of these components.  Thus, for the remainder of this paper, we
will focus on analyzing the harmonics polarized along the $x$-axis.
 Therefore, we can approximate $I_q = \xi_x \xi_x^* + \xi_y \xi_y^* \approx \xi_x
\xi_x^*$.

Moreover, the phase of the nonlinear polarization depends on the
intensity of the driving radiation \cite{LewensteinPhase,
ShinPhaseIntensity} and its ellipticity \cite{Antoine,Strelkov}. We
will ignore the effect of ellipticity on the phase of
$\vec{F}(I,\epsilon)$ since, as shown below, harmonic generation is
dominated by those regions in which the driving radiation is close
to linear polarization. We may write the intensity-dependent phase
as a Taylor expansion around $I_0$,
\begin{eqnarray}
\Phi(I) &= \Phi(I_0) + \left.\frac{d \Phi}{d I}\right| _{I_0} \left(I - I_0\right) + \ldots\\
&\approx \Phi_0 + \nu q \left(I - I_0\right).
\end{eqnarray}
where $\nu  = \left.d \Phi / d I \right| _{I_0}$. For simulations in
this paper, we assume $\nu/q \approx 0.2$ $\mathrm{rad}/
10^{14}$$\mathrm{W\,cm}^{-2}$ based on previous calculations
\cite{ShinPhaseIntensity}.

We may now gather the contributions to the total phase $\Psi_x(z)$:
\begin{eqnarray}
\Psi_x(z) &=& -\Phi(I_0) - \Delta k_1 z + q\psi_x(z) -  q\nu  \left(I - I_0\right)\\
&\approx& \Psi_0 - \Delta k_1 z - q\frac{c_2 M_2 }{c_1 M_1}\cos
\Omega \sin(\Delta \beta z)  \\&&- 2q \nu c_1M_1c_2M_2 \cos \Omega
\cos \left(\Delta \beta z\right) \label{eqn:phi simplified}
\end{eqnarray}
where $\Psi_0 = -\Phi(I_0) - 2 q\nu  c_1 M_1 c_2 M_2 \cos\Omega$,
and the approximation holds if $\left|c_2\right| \ll
\left|c_1\right|$.  From Eqns. \eqref{eqn: diffeq1} and
\eqref{eqn:phi simplified}, we can rewrite the the differential
equation for the $x$-component as:
\begin{eqnarray} \frac{d \xi_x}{dz} &=& \cos[\Theta(z)] \left| F[I(z), \varepsilon(z)] \right| e^{+i\Psi_x(z)} \\
&=& A' \Gamma(z) e^{-i\Psi_x'(z)} \label{eqn:xfinaldiffeq}
\end{eqnarray} where
\begin{eqnarray} A'(z) &=& A e^{+i \Psi_0} \\
\Gamma(z) &=&  \cos[\Theta(z)] \left| F[I(z), \varepsilon(z)] \right| \\
 \Psi_x'(z) &=& -\Psi_x(z) + \Psi_0  \\
&=& \Delta k_1 z + \gamma \sin(\Delta \beta z) + \rho \cos(\Delta
\beta z) \label{eqn:phasephase}
\end{eqnarray} in which $\gamma = q\frac{c_2 M_2 }{c_1 M_1}\cos \Omega$ is the mode interference phase term
and $\rho = 2q \nu c_1M_1c_2M_2 \cos \Omega$ is the
intensity-dependent phase term.

\section{Analysis of the Growth Equation \label{sec:fourier}}

\subsection{Phase Analysis using the Jacobi-Anger Expansion}
The exponential term in Eqn \eqref{eqn:xfinaldiffeq} can be expanded
into the products of two infinite sums using the Jacobi-Anger
Expansion:
\begin{eqnarray} e^{-i \Psi_x'(z)} &=& e^{-i \Delta k_1 z} \left[
\sum_{l=-\infty }^{+\infty} i^l J_l(-\rho) e^{+i l \Delta \beta z}
\right] \nonumber \\
&& \times \left[ \sum_{j=-\infty }^{+\infty} J_j(-\gamma) e^{+i j
\Delta \beta z} \right] \label{eqn:JAEFirst}
\end{eqnarray} where $J_l$ and $J_j$ are Bessel functions of the
first kind and $l, j \in \mathbb{Z}$. It is insightful to factor
terms of constant $\sigma = l + j$ to give:
\begin{eqnarray} e^{-i \Psi_x'(z)} &=& e^{ -i \Delta k_1 z}  \sum_{\sigma =-
\infty}^{\infty} U_\sigma e^{i \sigma \Delta \beta z}
\label{eqn:psisigma} \end{eqnarray} where
\begin{eqnarray} U_\sigma = \sum_{ l+j = \sigma} i^l J_l(-\rho)
J_j(-\gamma).\end{eqnarray}  We see that the modulation caused by
intensity dependent phase and mode beating can be resolved into
harmonics $\sigma \Delta \beta$ of the difference in spatial
frequency $\Delta \beta$ of the two modes.

\subsection{Source Amplitude Spatial Fourier Analysis}
The analysis above suggests that it would be useful to write the
source modulus $\Gamma$ as a superposition of Fourier components
with frequency $\kappa \Delta \beta$ (with $\kappa \in \mathbb{Z}$).
For periodic modulation of the driving radiation, the source modulus
can be written as a Fourier series:
\begin{eqnarray} \Gamma(z) = \sum_{\kappa = -\infty}^{\infty} V_\kappa e^{i \kappa \Delta \beta z} \end{eqnarray}
and hence the growth differential equation can be written as:
\begin{equation} \frac{1}{A'} \frac{d \xi_x}{d z} = e^{ -i \Delta k_1 z}
\sum_{\sigma, \kappa \in \mathbb{Z}} U_\sigma V_\kappa e^{i (\sigma
+ \kappa) \Delta \beta z} .\label{eqn:diffeqproduct}\end{equation}

The terms that contribute to monotonic harmonic growth are those for
which the phase is stationary, in other words, $\frac{\partial
\Psi_x}{\partial z} = 0$.  This implies that for QPM we require
$\Delta k_1 = (\sigma + \kappa) \Delta \beta$. We see that the
harmonics of the modulation frequency allow QPM of larger wave
vector mismatch $\Delta k$ or, equivalently, of shorter coherence
lengths $L_{c,1}$. The fundamental modulation spatial frequency
$\Delta \beta$ has a period $2L_b = 2 \pi / \Delta \beta$, and hence
we may write the QPM condition as $L_b = (\sigma + \kappa) L_{c,1} =
n L_{c,1}$, where $n = \sigma + \kappa$ is the order of the QPM
process. Factoring all the terms contributing to monotonic harmonic
growth, and ignoring the oscillating terms, the growth equation
becomes:
\begin{eqnarray} \frac{1}{A'} \frac{d \xi_x}{d z} &\approx & \sum_{\sigma + \kappa = n} U_{\sigma}V_{\kappa}
\label{eqn:difenvenv}
\end{eqnarray}
for a fixed $n$, keeping in mind that each of the terms of the sum
is complex and may have different signs. Eqn \eqref{eqn:difenvenv}
can easily be solved, from which the harmonic intensity is found to
be:
\begin{eqnarray} I_q \approx \frac{1}{2}A' S z^2 \label{eqn:envenv}\end{eqnarray}
where $S = \sum_{\sigma + \kappa = n} U_{\sigma}V_{\kappa}$. It is
useful to note that the $\sigma$ and $\kappa$ terms result from
phase and intensity modulation of the driver respectively.

\section{Simulation Results \label{sec:simulations}}

\subsection{Detailed simulations for $L_b = 2L_{c,1}$ and $L_b = L_{c,1}$}
\begin{figure}[tp]
\centering
\includegraphics[width=8.5cm]{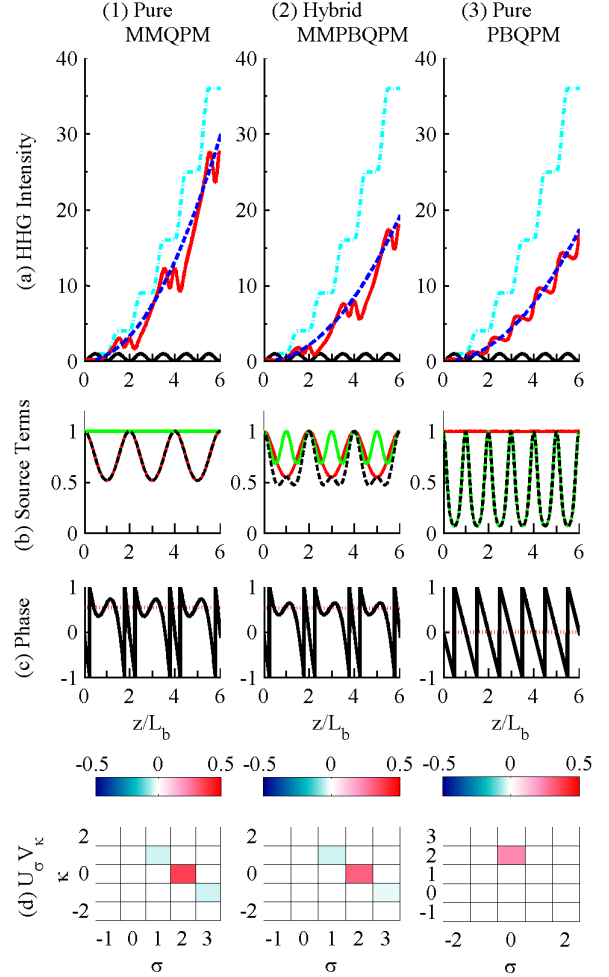}
\caption{Simulation results for Columns - (1) \lq \lq Pure MMQPM"
($c_2 M_2 = 0.002$, $\Omega = 0^o$) (2) \lq \lq Hybrid MMPBQPM"
($c_2 M_2 = 0.01$, $\Omega = 60^o$), and (3) \lq \lq Pure PBQPM"
($c_2 M_2 = 0.05$, $\Omega = 90^o$); all for $L_b = 2L_c$ and
$q=51$.  Row (a) shows the relative HHG intensity. Dot dashed cyan
line shows ideal square wave QPM, solid red line shows the indicated
form of MMPBQPM, solid black line shows the intensity for no phase
matching, and the dashed blue line shows the shows the harmonic
intensity calculated from Eqn \eqref{eqn:envenv}. Row (b) shows the
modulus of the intensity-dependent source term $G[I(z)]$ (solid red
line), ellipticity-dependent source term $H[\varepsilon(z)]$ (solid
green line), and the combined source terms $|F| = G H$ (dashed black
line). Row (c) shows $\Psi_x(z)$, the total phase of $d \xi_x/dz$
(solid black line) and $\Psi_x$ for large $z$ (dotted light red
line). Row (d) shows the $U_{\sigma} V_{\kappa}$ distribution as a
function of $\sigma$ and $\kappa$ for $n = \sigma + \kappa$ for
$n=2$ \label{fig:HHGGainx9}}
\end{figure}

\begin{figure}[tp]
\centering
\includegraphics[width=8.5cm]{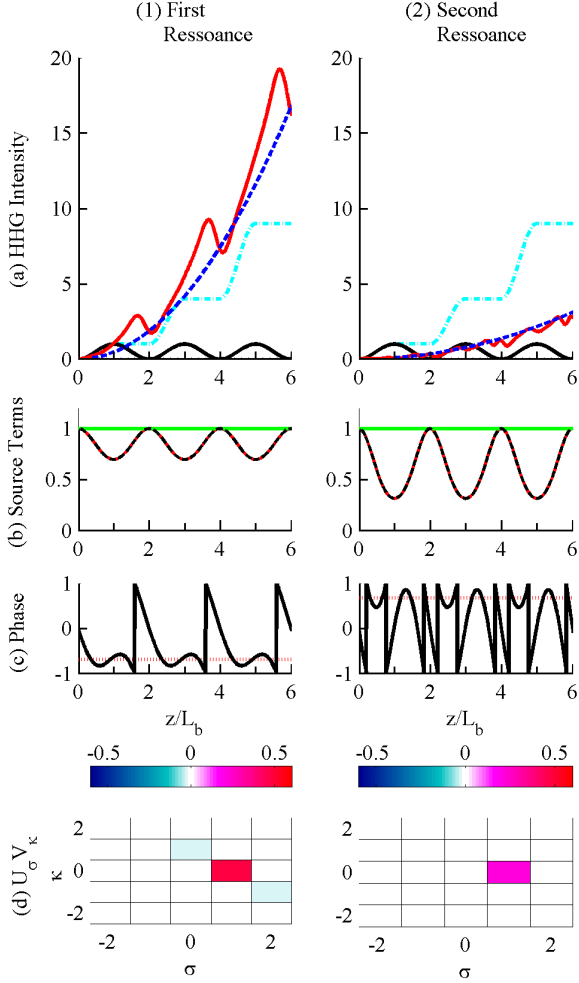}
\caption{Simulation results for Columns - (1) the first resonance
($c_2 M_2 = 0.0009$, $\Omega = 0^o$) and (2) the second resonance
($c_2 M_2 = 0.0091$, $\Omega = 0^o$); all for $L_b = L_c$ and
$q=51$.   Row (a) shows the relative HHG intensity. Dot dashed cyan
line shows ideal square wave QPM, solid red line shows the indicated
form of MMPBQPM, solid black line shows the intensity for no phase
matching, and the dashed blue line shows the shows the harmonic
intensity calculated from Eqn \eqref{eqn:envenv}. Row (b) shows the
modulus of the intensity-dependent source term $G[I(z)]$ (solid red
line), ellipticity-dependent source term $H[\varepsilon(z)]$ (solid
green line), and the combined source terms $|F| = G H$ (dashed black
line). Row (c) shows $\Psi_x(z)$, the total phase of $d \xi_x/dz$
(solid black line) and $\Psi_x$ for large $z$ (dotted light red
line). Row (d) shows the $U_{\sigma} V_{\kappa}$ distribution as a
function of $\sigma$ and $\kappa$ for $n = \sigma + \kappa$ for
$n=1$. \label{fig:HHGGainx8lclb}}
\end{figure}

To test these ideas, we have conducted a series of simulations. Fig.
\ref{fig:HHGGainx9}  presents the results of simulations for $L_b =
2 L_{c,1}$ ($n=2$) and three values of $\Omega$ and $c_2M_2$ for
$q=51$ while Fig.\ref{fig:HHGGainx8lclb} presents the same
parameters for $L_b = L_{c,1}$ ($n=1$), $\Omega = 0^o$ and two
different values of $c_2 M_2$ for $q=51$.  These values are compared
against ideal QPM, which is defined by the square wave modulation
between zero and one of the harmonic generation with a period of
$2nL_{c,1}$.

When $\Omega = 0^o$ MMPBQPM is equivalent to \lq \lq pure MMQPM"
since the driving radiation remains linearly polarized at all points
within the waveguide; this is seen in  Fig.\ref{fig:HHGGainx9}-Col
(1) where Fig.\ref{fig:HHGGainx9}-(1)(b) indicates that modulation
of the source term arises from mode-beating alone.

When $\Omega = 90^o$, MMPBQPM is equivalent to PBQPM since the mode
beating causes the polarization of the driving radiation to beat in
an analogous way to PBQPM driven by a linearly polarized beam
propagating in a birefringent waveguide.  This is seen in Fig.
\ref{fig:HHGGainx9}-Col (3).  More specifically, as seen in Fig.
\ref{fig:HHGGainx9}- (3)(b), the modulation of the source term is
seen to arise from solely polarization beating. It should be noted
that for $\Omega = 90^o$ the simulations presented here agree with
earlier calculations of PBQPM \cite{LiuPRAPBQPM}. For intermediate
values of $\Omega$ (such as in Col (2) where $\Omega = 60^o$),
modulation of both the intensity and polarization of the driving
radiation play a role in QPM.

Fig. \ref{fig:HHGGainx9} also compares the growth of the calculated
harmonic intensity with the approximation of Eqn.
\eqref{eqn:envenv}. It can be seen that the approximation agrees
closely with the exact calculation, indicating clearly the dominant
role played by the terms for which $\sigma + \kappa = n =2$ as seen
in Fig. \ref{fig:HHGGainx9} - Row (a).

 Moreover, Fig.\ref{fig:HHGGainx9} - Row (d) maps the values of $U_\sigma V_\kappa$, modulus phase (the terms in the sum in Eqn
\eqref{eqn:difenvenv}), as a function of $\sigma$ and $\kappa$ where
$\sigma + \kappa = n$ for a fixed $n=2$. Hence, only where $\sigma +
\kappa = 2$ is $U_\sigma V_\kappa$ is nonzero. For the case of Pure
MMQPM, Col (1),  and Hybrid MMPBQPM, Col (2), the dominant
contributing term is $(\sigma, \kappa) = (2, 0)$ indicating that QPM
arises predominantly from phase modulation of the driver, not
intensity modulation. This can also be seen in Fig.
\ref{fig:HHGGainx9}-1c and Fig. \ref{fig:HHGGainx9}-2c where the
regions of harmonic growth occur for points where the phase $\Psi_x$
is within $\pi/2$ of the phase of $\xi_x$ for large $z$. In
contrast, for the case of Pure PBQPM, Fig. \ref{fig:HHGGainx9} -
Col(3), the dominant term is $(\sigma, \kappa) = (0, 2)$. This
suggests, as expected, that for Pure PBQPM, phase modulation does
not contribute to QPM but only the modulation of the amplitude of
the source term caused by caused by polarization beating.

Fig. \ref{fig:HHGGainx8lclb} presents the same parameters in Fig.
\ref{fig:HHGGainx9} for $n=1$ (or $L_b = L_{c,1}$), for two
different values of $c_2M_2$ and $\Omega = 0^0$.  We see that for
both columns, the only terms which contribute are those for which
$\sigma + \kappa =1$, as expected.  For Col (1), optimal MMQPM
enables harmonics to be generated with intensities greater than for
ideal square wave QPM.  As indicated in Fig.
\ref{fig:HHGGainx8lclb}-(1)(c), the region of harmonic growth
coincides with $\Psi_x$ being within $\pm \pi/2$ of the phase of
$\xi_x$ for large $z$ .  Moreover, the largest contributing term of
$U_\sigma V_\kappa$ in Fig. \ref{fig:HHGGainx8lclb}-(1)(d) is
$(\sigma, \kappa) = (1, 0)$; this suggests that QPM is caused
primarily by phase modulation, and not by amplitude modulation as
reported earlier for MMQPM \cite{DromneyMMQPM, RobinsonThesis,
Zepf2007}. Moreover, the phase modulation explains why higher growth
than ideal square-wave QPM occurs. Fig. \ref{fig:HHGGainx8lclb} -
Col (2) shows the output at a different mode mix where $c_2 M_2 =
0.0091$ and $c_1 M_1 = 0.9909$. As discussed below, the mode
mixtures for which results are shown in Fig \ref{fig:HHGGainx8lclb}
correspond to two of the peaks in a plot of the output of harmonic
$q=51$ as a function of $c_1M_1$ and $c_2M_2$.

\subsection{Parameter Space Scans}
This section presents a series of parameter space scans for
optimizing the harmonic generation by \nobreak{MMPBQPM}. In an HHG
experiment, pressure, coupling angle $\Omega$, and the mode mix
ratio of $c_1$ to $c_2$ are parameters that can be adjusted.
Pressure tuning equates to tuning the coherence length, or tuning
the ratio $L_b/L_{c,1}$ assuming that $L_b$ is fixed for a specific
pair of driving and modifying modes.

\begin{figure}[tb]
\centering
\includegraphics[width=9.5cm]{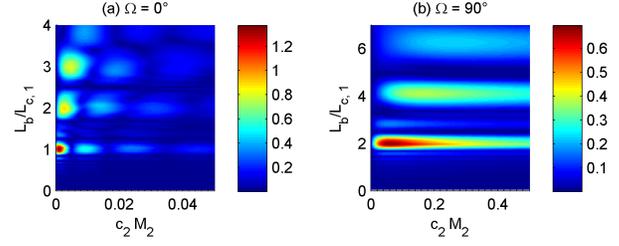}
\caption{Relative HHG amplitude $| \xi_x(z) |$ for $q=51$ at large
$z$ ($z=10L_{c,1}$) as a function of $L_b/L_{c,1}$ and $m=2$ mode
coupling strength $c_2^2 M_2^2$ where $c_1^2M_1^2  = 1- c_2^2M_2^2$,
normalized to ideal square wave QPM for: (a) $\Omega = 0^o$, and (b)
$\Omega = 90^o$ . \label{fig:PressureC2Scan}}
\end{figure}
Fig. \ref{fig:PressureC2Scan}a shows, for the MMQPM case ($\Omega =
0^o$), the variation of the harmonic output as a function of $
L_b/L_{c,1}$ and mode mix $c_2 M_2$. Note that here the magnitude of
the harmonic amplitude, not intensity, is plotted in order to show
more clearly the variation of the harmonic output.  As expected,
MMQPM is optimized for integer $n$.  Moreover, the peaks shift to
increasing $c_2M_2$ with increasing $n$.  When $n=1$, the QPM
condition becomes $\sigma + \kappa = l + j + \kappa = 1$. The three
lowest-order solutions satisfying this condition are $\{l,j,\kappa\}
= \{1,0,0\}, \{0,1,0\},$ and $\{0,0,1\}$.  We therefore expect peaks
in the HHG intensity to occur for values of $c_2M_2$ corresponding
to peaks in $|J_{1}(-\rho) J_{0} (-\gamma)|$, $|J_{0}(-\rho) J_{1}
(-\gamma)|$, or $|J_{0}(-\rho) J_{0} (-\gamma)|$. The maxima along
the line $L_b/L_{c,1}$ shown in Fig \ref{fig:PressureC2Scan}a arise
from the variations of $c_2M_2$ which optimize the functions of
$|J_l(-\rho) J_j(-\gamma)|$.

When $n = 2$, $\sigma + \kappa = l + j + \kappa =2$, and the three
lowest order terms are $\{1, 1,  0\}$, $\{0, 1, 1\}$, or $\{1, 0,
1\}$. Thus the values of optimal for $c_2 M_2$ will be around the
 extrema of $|J_{1}(-\rho) J_{1} (-\gamma)|$, $|J_{0}(-\rho) J_{1} (-\gamma)|$, $|J_{1}(-\rho) J_{0} (-\gamma)|$.
 Because the positions of the local extremas of the Bessel function increase with increasing $|l|$ or $|j|$,
 optimal $\rho$ and $\gamma$ and thus optimal $c_2 M_2$ will increase as well.  Therefore, increasing $n$ will result in
 larger values of $|l|$ and $|j|$ contributing to the harmonic
 growth corresponding to the Bessel function peaks shifted to higher
 values of $(-\rho)$ and $(-\gamma)$ and hence higher values of $c_2 M_2$.

\begin{figure}[tb]
\centering
\includegraphics[width=9cm]{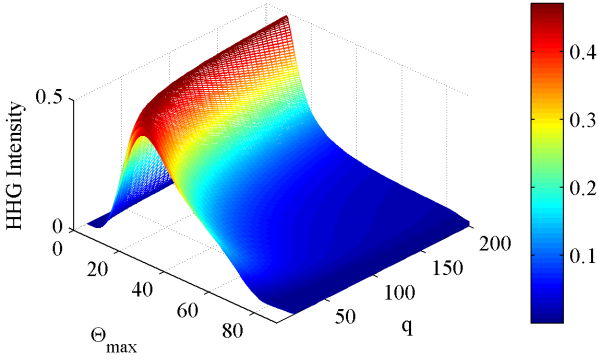}
\caption{Relative HHG intensity at $z=L_b$ for $L_b = 2L_{c,1}$ as a
function of $\Theta_{max} = \tan^{-1} \left(\frac{c_2 M_2}{c_1
M_1}\right)$ and $q$, normalized to ideal QPM.
\label{fig:PBQPMParameters}}
\end{figure}

\begin{figure}[tb]
\centering
\includegraphics[width=9cm]{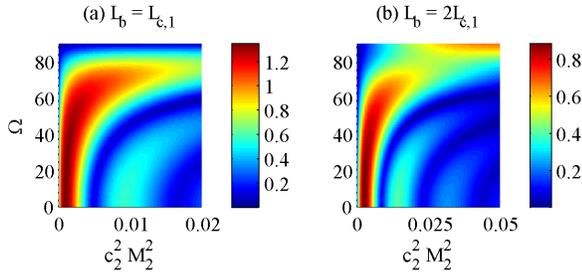}
\caption{Relative HHG amplitude for $q=51$ after $2L_b$ as a
function of coupling angle $\Omega$ and $m=2$ mode coupling strength
$c_2^2 M_2^2$ where $c_1^2M_1^2  = 1- c_2^2M_2^2$, normalized to
ideal QPM for: (a) $L_b = L_{c,1}$, and (b) $L_b = 2 L_{c,1}$.
\label{fig:Parameters}}
\end{figure}

Similarly, Fig. \ref{fig:PressureC2Scan}b shows the PBQPM case where
$\Omega = 90^0$.  As discussed in \cite{LiuPRAPBQPM}, PBQPM will
occur when $L_b = n L_{c,1}$ and $n$ is even -- as is evident  in
Fig. \ref{fig:PressureC2Scan}b. Since $\Omega = 90^o$, $\rho =
\gamma = 0$, and since $J_l(0)J_j(0) = 0$ unless $l=j=0$, monotonic
harmonic growth can only occur for $\sigma = l + k = 0$. Hence
optimal PBQPM occurs when the Fourier coefficient $V_k$ is large for
even $\kappa$.  Furthermore, PBQPM does not contribute to any phase
modulation because $\sigma = 0$ as seen from Eqn
\eqref{eqn:psisigma}.  The optimal value of $c_2M_2$ increases with
the order $n$ of QPM since increasing this parameter shifts the
Fourier spectrum of the driving intensity modulations to higher
orders.  The optimal value of $c_2M_2$ is explored more clearly in
Fig. \ref{fig:PBQPMParameters}, which shows the normalized harmonic
intensity for $\Omega = 90^o$ as a function of the harmonic  order q
and the maximum angle $\Theta_{max}= \tan^{-1} \left(
\frac{c_2M_2}{c_1M_1} \right)$ the major axis of the elliptical
driving radiation makes with the $x$-axis. If $\Theta_{max}$ is too
close to $0^o$, then the ellipticity modulation is not enough to
suppress the destructive zones.  If $\Theta_{max}$ is too close to
$90^o$, then the harmonic generation suppression zone is too large
to create efficient harmonics.

Fig. \ref{fig:Parameters} plots, for the cases $L_b = L_{c,1}$ and
$L_b = 2 L_{c,1}$,
 the calculated relative amplitude at $z = 2L_b$ of the $q = 51$ harmonic as a function of $\Omega$ and the relative intensity of the $m= 2$ mode.
In the case of Fig. \ref{fig:Parameters}a, $L_b = L_{c,1}$, the
relative amplitude achieved with \lq \lq pure MMQPM" (i.e. $\Omega
=0$) is greater than that of ideal square wave QPM as explained
above. For $\Omega = 0^o$, the intensity oscillates with increasing
$c_2 M_2$, with the size of the resonant peaks decreasing with
increasing $c_2 M_2$. These resonance peaks are caused by peaks of
the products $J_l(-\rho) J_j(-\gamma)$ with $\rho$ and $\gamma$
being a linear function of $c_2 M_2$ (for small $c_2M_2$) as
explained above. For the case of pure PBQPM, i.e. $\Omega = 90^o$
and $L_b = L_{c,1}$, the harmonic intensity is seen to be very low
and almost independent of the relative intensity of the two modes
since the phase-matching condition for lowest-order PBQPM is not
satisfied, and QPM is not achieved.

Lowest-order QPM occurs for $n=2$, as shown in Fig
\ref{fig:Parameters}b, a region of bright harmonic generation occurs
for $\Omega = 90^o$ and $c_2M_2 \approx 0.04$. When $\Omega = 0^o$
the harmonic intensity  oscillates in a similar manner to that
observed in Fig. \ref{fig:Parameters}a.

\section{Conclusion}
We have developed a generalized analysis of MMQPM and PBQPM together
with a simplified Fourier analysis which gives additional insights
into the dominant contributions of quasi phase-matching.  In
addition we have shown that PBQPM could be achieved without a
birefringent waveguide by exciting a pair of waveguide modes with
two orthogonal polarizations.

Our analysis of MMQPM showed, in contrast to our earlier analysis
\cite{DromneyMMQPM, Zepf2007}, that QPM is dominated by the
modulation of phase of the harmonic source term, not of its
amplitude.  This allows, under optimal conditions, MMQPM to generate
harmonics with an intensity greater than possible with ideal,
square-wave QPM.

The authors would like to thank the EPRSC for support through grant
No. EP/GO67694/1. Lewis Liu would like thank David Lloyd for
fruitful discussions and the James Buckee Scholarship of Merton
College, Oxford for its support.

\section{APS Copyright Notice}
Copyright to the [above-listed] unpublished and original article submitted by the [above] author(s), the abstract forming part thereof, and any subsequent errata (collectively, the “Article”) is hereby transferred to the American Physical Society (APS) for the full term thereof throughout the world, subject to the Author Rights (as hereinafter defined) and to acceptance of the Article for publication in a journal of APS. This transfer of copyright includes all material to be published as part of the Article (in any medium), including but not limited to tables, figures, graphs, movies, other multimedia files, and all supplemental materials. APS shall have the right to register copyright to the Article in its name as claimant, whether separately or as part of the journal issue or other medium in which the Article is included.

\bibliography{MMPBQPMHHG_BibV7}

\begin{thebibliography}{10}%
\makeatletter
\providecommand \@ifxundefined [1]{%
 \ifx #1\undefined \expandafter \@firstoftwo
 \else \expandafter \@secondoftwo
\fi
}%
\providecommand \@ifnum [1]{%
 \ifnum #1\expandafter \@firstoftwo
 \else \expandafter \@secondoftwo
\fi
}%
\providecommand \enquote [1]{``#1''}%
\providecommand \bibnamefont  [1]{#1}%
\providecommand \bibfnamefont [1]{#1}%
\providecommand \citenamefont [1]{#1}%
\providecommand\href[0]{\@sanitize\@href}%
\providecommand\@href[1]{\endgroup\@@startlink{#1}\endgroup\@@href}%
\providecommand\@@href[1]{#1\@@endlink}%
\providecommand \@sanitize [0]{\begingroup\catcode`\&12\catcode`\#12\relax}%
\@ifxundefined \pdfoutput {\@firstoftwo}{%
 \@ifnum{\z@=\pdfoutput}{\@firstoftwo}{\@secondoftwo}%
}{%
 \providecommand\@@startlink[1]{\leavevmode\special{html:<a href="#1">}}%
 \providecommand\@@endlink[0]{\special{html:</a>}}%
}{%
 \providecommand\@@startlink[1]{%
  \leavevmode
  \pdfstartlink
   attr{/Border[0 0 1 ]/H/I/C[0 1 1]}%
   user{/Subtype/Link/A<</Type/Action/S/URI/URI(#1)>>}%
  \relax
 }%
 \providecommand\@@endlink[0]{\pdfendlink}%
}%
\providecommand \url  [0]{\begingroup\@sanitize \@url }%
\providecommand \@url [1]{\endgroup\@href {#1}{\urlprefix}}%
\providecommand \urlprefix [0]{URL }%
\providecommand \Eprint[0]{\href }%
\@ifxundefined \urlstyle {%
  \providecommand \doi [1]{doi:\discretionary{}{}{}#1}%
}{%
  \providecommand \doi [0]{doi:\discretionary{}{}{}\begingroup
  \urlstyle{rm}\Url }%
}%
\providecommand \doibase [0]{http://dx.doi.org/}%
\providecommand \Doi[1]{\href{\doibase#1}}%
\providecommand \bibAnnote [3]{%
  \BibitemShut{#1}%
  \begin{quotation}\noindent
    \textsc{Key:}\ #2\\\textsc{Annotation:}\ #3%
  \end{quotation}%
}%
\providecommand \bibAnnoteFile [2]{%
  \IfFileExists{#2}{\bibAnnote {#1} {#2} {\input{#2}}}{}%
}%
\providecommand \typeout [0]{\immediate \write \m@ne }%
\providecommand \selectlanguage [0]{\@gobble}%
\providecommand \bibinfo [0]{\@secondoftwo}%
\providecommand \bibfield [0]{\@secondoftwo}%
\providecommand \translation [1]{[#1]}%
\providecommand \BibitemOpen[0]{}%
\providecommand \bibitemStop [0]{}%
\providecommand \bibitemNoStop [0]{.\EOS\space}%
\providecommand \EOS [0]{\spacefactor3000\relax}%
\providecommand \BibitemShut [1]{\csname bibitem#1\endcsname}%
\bibitem{Uiberacker}%
  \BibitemOpen
  \bibfield{author}{%
  \bibinfo {author} {\bibfnamefont{M.}~\bibnamefont{Uiberacker}}\ and\ \bibinfo
  {author} {\bibnamefont{et~al}},\ }%
  \bibfield{journal}{%
  \bibinfo {journal} {Nature}\ }%
  \textbf{\bibinfo {volume} {446}},\ \bibinfo {pages} {627} (\bibinfo {year}
  {2007})%
  \bibAnnoteFile{NoStop}{Uiberacker}%
\bibitem{Schultze}%
  \BibitemOpen
  \bibfield{author}{%
  \bibinfo {author} {\bibfnamefont{M.}~\bibnamefont{Schultze}}\ and\ \bibinfo
  {author} {\bibnamefont{et~al}},\ }%
  \bibfield{journal}{%
  \bibinfo {journal} {Science}\ }%
  \textbf{\bibinfo {volume} {328}},\ \bibinfo {pages} {1658} (\bibinfo {year}
  {2010})%
  \bibAnnoteFile{NoStop}{Schultze}%
\bibitem{Cavalieri}%
  \BibitemOpen
  \bibfield{author}{%
  \bibinfo {author} {\bibfnamefont{A.}~\bibnamefont{Cavalieri}}\ and\ \bibinfo
  {author} {\bibnamefont{et~al}},\ }%
  \bibfield{journal}{%
  \bibinfo {journal} {Nature}\ }%
  \textbf{\bibinfo {volume} {449}},\ \bibinfo {pages} {1029} (\bibinfo {year}
  {2007})%
  \bibAnnoteFile{NoStop}{Cavalieri}%
\bibitem{RaHolography}%
  \BibitemOpen
  \bibfield{author}{%
  \bibinfo {author} {\bibfnamefont{R.~I.}\ \bibnamefont{Tobey}}, \bibinfo
  {author} {\bibfnamefont{M.~E.}\ \bibnamefont{Siemens}}, \bibinfo {author}
  {\bibfnamefont{O.}~\bibnamefont{Cohen}}, \bibinfo {author}
  {\bibfnamefont{M.~M.}\ \bibnamefont{Murnane}}, \bibinfo {author}
  {\bibfnamefont{H.~C.}\ \bibnamefont{Kapteyn}},\ and\ \bibinfo {author}
  {\bibfnamefont{K.~A.}\ \bibnamefont{Nelson}},\ }%
  \bibfield{journal}{%
  \bibinfo {journal} {Optics Letters}\ }%
  \textbf{\bibinfo {volume} {32(3)}},\ \bibinfo {pages} {286} (\bibinfo {year}
  {2007})%
  \bibAnnoteFile{NoStop}{RaHolography}%
\bibitem{Sandberg}%
  \BibitemOpen
  \bibfield{author}{%
  \bibinfo {author} {\bibfnamefont{R.~L.}\ \bibnamefont{Sandberg}}\ and\
  \bibinfo {author} {\bibnamefont{et~al}},\ }%
  \bibfield{journal}{%
  \bibinfo {journal} {Phys. Rev. Lett.}\ }%
  \textbf{\bibinfo {volume} {99}},\ \bibinfo {pages} {098103} (\bibinfo {year}
  {2007})%
  \bibAnnoteFile{NoStop}{Sandberg}%
\bibitem{Corkum}%
  \BibitemOpen
  \bibfield{author}{%
  \bibinfo {author} {\bibfnamefont{P.~B.}\ \bibnamefont{Corkum}},\ }%
  \bibfield{journal}{%
  \bibinfo {journal} {Phys. Rev. Lett.}\ }%
  \textbf{\bibinfo {volume} {71(13)}},\ \bibinfo {pages} {1994} (\bibinfo
  {year} {1993})%
  \bibAnnoteFile{NoStop}{Corkum}%
\bibitem{Lewenstien}%
  \BibitemOpen
  \bibfield{author}{%
  \bibinfo {author} {\bibfnamefont{M.}~\bibnamefont{Lewenstein}}, \bibinfo
  {author} {\bibfnamefont{P.}~\bibnamefont{Yu}}, \bibinfo {author}
  {\bibfnamefont{A.}~\bibnamefont{L'Huillier}},\ and\ \bibinfo {author}
  {\bibfnamefont{P.~B.}\ \bibnamefont{Corkum}},\ }%
  \bibfield{journal}{%
  \bibinfo {journal} {Physical Review A}\ }%
  \textbf{\bibinfo {volume} {49(3)}},\ \bibinfo {pages} {2117} (\bibinfo {year}
  {1994})%
  \bibAnnoteFile{NoStop}{Lewenstien}%
\bibitem{Murnane1999}%
  \BibitemOpen
  \bibfield{author}{%
  \bibinfo {author} {\bibfnamefont{C.~G.}\ \bibnamefont{Durfee}}, \bibinfo
  {author} {\bibfnamefont{A.~R.}\ \bibnamefont{Rundquist}}, \bibinfo {author}
  {\bibfnamefont{S.}~\bibnamefont{Backus}}, \bibinfo {author}
  {\bibfnamefont{C.}~\bibnamefont{Herne}}, \bibinfo {author}
  {\bibfnamefont{M.~M.}\ \bibnamefont{Murnane}},\ and\ \bibinfo {author}
  {\bibfnamefont{H.~C.}\ \bibnamefont{Kapteyn}},\ }%
  \bibfield{journal}{%
  \bibinfo {journal} {Phys. Rev. Lett.}\ }%
  \textbf{\bibinfo {volume} {83(11)}},\ \bibinfo {pages} {2187} (\bibinfo
  {year} {1999})%
  \bibAnnoteFile{NoStop}{Murnane1999}%
\bibitem{Murnane1998}%
  \BibitemOpen
  \bibfield{author}{%
  \bibinfo {author} {\bibfnamefont{A.~R.}\ \bibnamefont{Rundquist}}, \bibinfo
  {author} {\bibfnamefont{C.~G.}\ \bibnamefont{Durfee}}, \bibinfo {author}
  {\bibfnamefont{Z.}~\bibnamefont{Chang}}, \bibinfo {author}
  {\bibfnamefont{S.~B.}\ \bibnamefont{C.~Herne}}, \bibinfo {author}
  {\bibfnamefont{M.~M.}\ \bibnamefont{Murnane}},\ and\ \bibinfo {author}
  {\bibfnamefont{H.~C.}\ \bibnamefont{Kapteyn}},\ }%
  \bibfield{journal}{%
  \bibinfo {journal} {Science}\ }%
  \textbf{\bibinfo {volume} {280(5368)}},\ \bibinfo {pages} {1412} (\bibinfo
  {year} {1998})%
  \bibAnnoteFile{NoStop}{Murnane1998}%
\bibitem{MurnaneScience2012}%
  \BibitemOpen
  \bibfield{author}{%
  \bibinfo {author} {\bibfnamefont{T.}~\bibnamefont{Popmintchev}}\ and\
  \bibinfo {author} {\bibnamefont{et~al}},\ }%
  \bibfield{journal}{%
  \bibinfo {journal} {Science}\ }%
  \textbf{\bibinfo {volume} {8}},\ \bibinfo {pages} {1287} (\bibinfo {year}
  {2012})%
  \bibAnnoteFile{NoStop}{MurnaneScience2012}%
\bibitem{Robinson2010}%
  \BibitemOpen
  \bibfield{author}{%
  \bibinfo {author} {\bibfnamefont{T.}~\bibnamefont{Robinson}}, \bibinfo
  {author} {\bibfnamefont{K.}~\bibnamefont{O'Keeffe}}, \bibinfo {author}
  {\bibfnamefont{M.}~\bibnamefont{Zepf}}, \bibinfo {author}
  {\bibfnamefont{B.}~\bibnamefont{Dromey}},\ and\ \bibinfo {author}
  {\bibfnamefont{S.~M.}\ \bibnamefont{Hooker}},\ }%
  \bibfield{journal}{%
  \bibinfo {journal} {J. Opt. Soc. Am. B}\ }%
  \textbf{\bibinfo {volume} {27}},\ \bibinfo {pages} {763} (\bibinfo {year}
  {2010})%
  \bibAnnoteFile{NoStop}{Robinson2010}%
\bibitem{Dromey}%
  \BibitemOpen
  \bibfield{author}{%
  \bibinfo {author} {\bibfnamefont{B.}~\bibnamefont{Dromey}}, \bibinfo {author}
  {\bibfnamefont{M.}~\bibnamefont{Zepf}}, \bibinfo {author}
  {\bibfnamefont{M.}~\bibnamefont{Landreman}}, \bibinfo {author}
  {\bibfnamefont{K.}~\bibnamefont{O'Keeffe}}, \bibinfo {author}
  {\bibfnamefont{T.}~\bibnamefont{Robinson}},\ and\ \bibinfo {author}
  {\bibfnamefont{S.~M.}\ \bibnamefont{Hooker}},\ }%
  \bibfield{journal}{%
  \bibinfo {journal} {Applied Optics}\ }%
  \textbf{\bibinfo {volume} {46}},\ \bibinfo {pages} {5142} (\bibinfo {year}
  {2007})%
  \bibAnnoteFile{NoStop}{Dromey}%
\bibitem{Peatross}%
  \BibitemOpen
  \bibfield{author}{%
  \bibinfo {author} {\bibfnamefont{J.}~\bibnamefont{Peatross}}, \bibinfo
  {author} {\bibfnamefont{S.}~\bibnamefont{Voronov}},\ and\ \bibinfo {author}
  {\bibfnamefont{I.}~\bibnamefont{Prokopovich}},\ }%
  \bibfield{journal}{%
  \bibinfo {journal} {Opt. Express}\ }%
  \textbf{\bibinfo {volume} {1(5)}},\ \bibinfo {pages} {114} (\bibinfo {year}
  {1997})%
  \bibAnnoteFile{NoStop}{Peatross}%
\bibitem{Lytle}%
  \BibitemOpen
  \bibfield{author}{%
  \bibinfo {author} {\bibfnamefont{A.~L.}\ \bibnamefont{Lytle}}, \bibinfo
  {author} {\bibfnamefont{X.}~\bibnamefont{Zhang}}, \bibinfo {author}
  {\bibfnamefont{P.}~\bibnamefont{Arpin}}, \bibinfo {author}
  {\bibfnamefont{O.}~\bibnamefont{Cohen}}, \bibinfo {author}
  {\bibfnamefont{M.~M.}\ \bibnamefont{Murnane}},\ and\ \bibinfo {author}
  {\bibfnamefont{H.~C.}\ \bibnamefont{Kapteyn}},\ }%
  \bibfield{journal}{%
  \bibinfo {journal} {Optics Letters}\ }%
  \textbf{\bibinfo {volume} {33(2)}},\ \bibinfo {pages} {174} (\bibinfo {year}
  {2008})%
  \bibAnnoteFile{NoStop}{Lytle}%
\bibitem{Zhang}%
  \BibitemOpen
  \bibfield{author}{%
  \bibinfo {author} {\bibfnamefont{X.}~\bibnamefont{Zhang}}, \bibinfo {author}
  {\bibfnamefont{A.~L.}\ \bibnamefont{Lytle}}, \bibinfo {author}
  {\bibfnamefont{T.}~\bibnamefont{Popmintchev}}, \bibinfo {author}
  {\bibfnamefont{X.}~\bibnamefont{Zhou}}, \bibinfo {author}
  {\bibfnamefont{H.~C.}\ \bibnamefont{Kapteyn}}, \bibinfo {author}
  {\bibfnamefont{M.~M.}\ \bibnamefont{Murnane}},\ and\ \bibinfo {author}
  {\bibfnamefont{O.}~\bibnamefont{Cohen}},\ }%
  \bibfield{journal}{%
  \bibinfo {journal} {Nature Physics}\ }%
  \textbf{\bibinfo {volume} {3(4)}},\ \bibinfo {pages} {270} (\bibinfo {year}
  {2007})%
  \bibAnnoteFile{NoStop}{Zhang}%
\bibitem{Zepf2007}%
  \BibitemOpen
  \bibfield{author}{%
  \bibinfo {author} {\bibfnamefont{M.}~\bibnamefont{Zepf}}, \bibinfo {author}
  {\bibfnamefont{B.}~\bibnamefont{Dromey}}, \bibinfo {author}
  {\bibfnamefont{M.}~\bibnamefont{Landreman}}, \bibinfo {author}
  {\bibfnamefont{P.}~\bibnamefont{Foster}},\ and\ \bibinfo {author}
  {\bibfnamefont{S.~M.}\ \bibnamefont{Hooker}},\ }%
  \bibfield{journal}{%
  \bibinfo {journal} {Phys. Rev. Lett.}\ }%
  \textbf{\bibinfo {volume} {99}},\ \bibinfo {pages} {143901} (\bibinfo {year}
  {2007})%
  \bibAnnoteFile{NoStop}{Zepf2007}%
\bibitem{RobinsonThesis}%
  \BibitemOpen
  \bibfield{author}{%
  \bibinfo {author} {\bibfnamefont{T.}~\bibnamefont{Robinson}},\ }%
  \emph{\bibinfo {title} {Quasi-Phase-Matching of High-Harmonic Generation}},\
  \bibinfo {type} {{D.Phil} thesis},\ \bibinfo {school} {University of Oxford}
  (\bibinfo {year} {2007})%
  \bibAnnoteFile{NoStop}{RobinsonThesis}%
\bibitem{DromneyMMQPM}%
  \BibitemOpen
  \bibfield{author}{%
  \bibinfo {author} {\bibfnamefont{B.}~\bibnamefont{Dromney}}\ and\ \bibinfo
  {author} {\bibnamefont{et~al}},\ }%
  \bibfield{journal}{%
  \bibinfo {journal} {Opt. Express}\ }%
  \textbf{\bibinfo {volume} {15(13)}},\ \bibinfo {pages} {7894} (\bibinfo
  {year} {2007})%
  \bibAnnoteFile{NoStop}{DromneyMMQPM}%
\bibitem{WalterMMQPM}%
  \BibitemOpen
  \bibfield{author}{%
  \bibinfo {author} {\bibfnamefont{D.}~\bibnamefont{Walter}}\ and\ \bibinfo
  {author} {\bibnamefont{et~al}},\ }%
  \bibfield{journal}{%
  \bibinfo {journal} {Opt. Express}\ }%
  \textbf{\bibinfo {volume} {14(6)}},\ \bibinfo {pages} {3433} (\bibinfo {year}
  {2006})%
  \bibAnnoteFile{NoStop}{WalterMMQPM}%
\bibitem{ModWaveguide}%
  \BibitemOpen
  \bibfield{author}{%
  \bibinfo {author} {\bibfnamefont{I.}~\bibnamefont{Christov}}, \bibinfo
  {author} {\bibfnamefont{H.}~\bibnamefont{Kapteyn}},\ and\ \bibinfo {author}
  {\bibfnamefont{M.}~\bibnamefont{Murnane}},\ }%
  \bibfield{journal}{%
  \bibinfo {journal} {Opt. Express}\ }%
  \textbf{\bibinfo {volume} {7(11)}},\ \bibinfo {pages} {362} (\bibinfo {year}
  {2000})%
  \bibAnnoteFile{NoStop}{ModWaveguide}%
\bibitem{LiuPRAPBQPM}%
  \BibitemOpen
  \bibfield{author}{%
  \bibinfo {author} {\bibfnamefont{L.~Z.}\ \bibnamefont{Liu}}, \bibinfo
  {author} {\bibfnamefont{K.}~\bibnamefont{O'Keeffe}},\ and\ \bibinfo {author}
  {\bibfnamefont{S.~M.}\ \bibnamefont{Hooker}},\ }%
  \bibfield{journal}{%
  \bibinfo {journal} {Phys Rev. A}\ }%
  \textbf{\bibinfo {volume} {85}},\ \bibinfo {pages} {053823} (\bibinfo {year}
  {2012})%
  \bibAnnoteFile{NoStop}{LiuPRAPBQPM}%
\bibitem{LiuOptLettORQPM}%
  \BibitemOpen
  \bibfield{author}{%
  \bibinfo {author} {\bibfnamefont{L.~Z.}\ \bibnamefont{Liu}}, \bibinfo
  {author} {\bibfnamefont{K.}~\bibnamefont{O'Keeffe}},\ and\ \bibinfo {author}
  {\bibfnamefont{S.~M.}\ \bibnamefont{Hooker}},\ }%
  \bibfield{journal}{%
  \bibinfo {journal} {Opt. Lett.}\ }%
  \textbf{\bibinfo {volume} {37(12)}},\ \bibinfo {pages} {2415} (\bibinfo
  {year} {2012})%
  \bibAnnoteFile{NoStop}{LiuOptLettORQPM}%
\bibitem{PatentPBQPM}%
  \BibitemOpen
  \bibfield{author}{%
  \bibinfo {author} {\bibfnamefont{L.~Z.}\ \bibnamefont{Liu}}, \bibinfo
  {author} {\bibfnamefont{K.}~\bibnamefont{O'Keeffe}},\ and\ \bibinfo {author}
  {\bibfnamefont{S.~M.}\ \bibnamefont{Hooker}},\ }%
  \enquote{\bibinfo {title} {High harmonic optical generator [polarization
  beating]},}\ \bibinfo {howpublished} {{Isis Innovation, U.K. Patent
  Application No.} GB1117355.6} (\bibinfo {year} {07~Oct.~2011})%
  \bibAnnoteFile{NoStop}{PatentPBQPM}%
\bibitem{PatentORQPM}%
  \BibitemOpen
  \bibfield{author}{%
  \bibinfo {author} {\bibfnamefont{L.~Z.}\ \bibnamefont{Liu}}, \bibinfo
  {author} {\bibfnamefont{K.}~\bibnamefont{O'Keeffe}},\ and\ \bibinfo {author}
  {\bibfnamefont{S.~M.}\ \bibnamefont{Hooker}},\ }%
  \enquote{\bibinfo {title} {High harmonic optical generato [optical
  rotation]},}\ \bibinfo {howpublished} {{Isis Innovation, U.K. Patent
  Application No.} GB1208753.2} (\bibinfo {year} {18~May~2012})%
  \bibAnnoteFile{NoStop}{PatentORQPM}%
\bibitem{Bahabad}%
  \BibitemOpen
  \bibfield{author}{%
  \bibinfo {author} {\bibfnamefont{A.}~\bibnamefont{Bahabad}}\ and\ \bibinfo
  {author} {\bibnamefont{et~al}},\ }%
  \bibfield{journal}{%
  \bibinfo {journal} {Nature Photonics}\ }%
  \textbf{\bibinfo {volume} {4}},\ \bibinfo {pages} {570} (\bibinfo {year}
  {2010})%
  \bibAnnoteFile{NoStop}{Bahabad}%
\bibitem{Shkolnikov}%
  \BibitemOpen
  \bibfield{author}{%
  \bibinfo {author} {\bibfnamefont{P.~L.}\ \bibnamefont{Shkolnikov}}, \bibinfo
  {author} {\bibfnamefont{A.~E.}\ \bibnamefont{Kaplan}},\ and\ \bibinfo
  {author} {\bibfnamefont{A.}~\bibnamefont{Lago}},\ }%
  \bibfield{journal}{%
  \bibinfo {journal} {J. Opt. Soc. Am. B}\ }%
  \textbf{\bibinfo {volume} {13(2)}},\ \bibinfo {pages} {412} (\bibinfo {year}
  {1996})%
  \bibAnnoteFile{NoStop}{Shkolnikov}%
\bibitem{Antoine}%
  \BibitemOpen
  \bibfield{author}{%
  \bibinfo {author} {\bibfnamefont{P.}~\bibnamefont{Antoine}}\ and\ \bibinfo
  {author} {\bibnamefont{et~al}},\ }%
  \bibfield{journal}{%
  \bibinfo {journal} {Phys. Rev. A}\ }%
  \textbf{\bibinfo {volume} {53(3)}},\ \bibinfo {pages} {1725} (\bibinfo {year}
  {1996})%
  \bibAnnoteFile{NoStop}{Antoine}%
\bibitem{Antoine2}%
  \BibitemOpen
  \bibfield{author}{%
  \bibinfo {author} {\bibfnamefont{P.}~\bibnamefont{Antoine}}\ and\ \bibinfo
  {author} {\bibnamefont{et~al}},\ }%
  \bibfield{journal}{%
  \bibinfo {journal} {Phys. Rev. A}\ }%
  \textbf{\bibinfo {volume} {55(2)}},\ \bibinfo {pages} {1314} (\bibinfo {year}
  {1997})%
  \bibAnnoteFile{NoStop}{Antoine2}%
\bibitem{Budil1993}%
  \BibitemOpen
  \bibfield{author}{%
  \bibinfo {author} {\bibfnamefont{K.~S.}\ \bibnamefont{Budil}}, \bibinfo
  {author} {\bibnamefont{P.Salieres}}, \bibinfo {author} {\bibfnamefont{M.~D.}\
  \bibnamefont{Perry}},\ and\ \bibinfo {author}
  {\bibfnamefont{A.}~\bibnamefont{L'huillier}},\ }%
  \bibfield{journal}{%
  \bibinfo {journal} {Phys. Rev. A}\ }%
  \textbf{\bibinfo {volume} {48(5)}},\ \bibinfo {pages} {R3437} (\bibinfo
  {year} {1993})%
  \bibAnnoteFile{NoStop}{Budil1993}%
\bibitem{DietrichPolarization}%
  \BibitemOpen
  \bibfield{author}{%
  \bibinfo {author} {\bibfnamefont{P.}~\bibnamefont{Dietrich}}\ and\ \bibinfo
  {author} {\bibnamefont{et~al}},\ }%
  \bibfield{journal}{%
  \bibinfo {journal} {Phys. Rev. A}\ }%
  \textbf{\bibinfo {volume} {50(5)}},\ \bibinfo {pages} {3585} (\bibinfo {year}
  {1994})%
  \bibAnnoteFile{NoStop}{DietrichPolarization}%
\bibitem{SchulzePolarization}%
  \BibitemOpen
  \bibfield{author}{%
  \bibinfo {author} {\bibfnamefont{D.}~\bibnamefont{Schulze}}\ and\ \bibinfo
  {author} {\bibnamefont{et~al}},\ }%
  \bibfield{journal}{%
  \bibinfo {journal} {Phys. Rev. A}\ }%
  \textbf{\bibinfo {volume} {57(4)}},\ \bibinfo {pages} {3003} (\bibinfo {year}
  {1998})%
  \bibAnnoteFile{NoStop}{SchulzePolarization}%
\bibitem{SolaPolarization}%
  \BibitemOpen
  \bibfield{author}{%
  \bibinfo {author} {\bibfnamefont{I.~J.}\ \bibnamefont{Sola}}\ and\ \bibinfo
  {author} {\bibnamefont{et~al}},\ }%
  \bibfield{journal}{%
  \bibinfo {journal} {Nature Physics}\ }%
  \textbf{\bibinfo {volume} {2}},\ \bibinfo {pages} {320} (\bibinfo {year}
  {2006})%
  \bibAnnoteFile{NoStop}{SolaPolarization}%
\bibitem{Soviet2}%
  \BibitemOpen
  \bibfield{author}{%
  \bibinfo {author} {\bibfnamefont{N.~L.}\ \bibnamefont{Manakov}}\ and\
  \bibinfo {author} {\bibfnamefont{Z.}~\bibnamefont{Ovsyannikov}},\ }%
  \bibfield{journal}{%
  \bibinfo {journal} {Sov. Phys. JETP}\ }%
  \textbf{\bibinfo {volume} {52}}%
  \bibAnnoteFile{NoStop}{Soviet2}%
\bibitem{Strelkov}%
  \BibitemOpen
  \bibfield{author}{%
  \bibinfo {author} {\bibfnamefont{V.}~\bibnamefont{Strelkov}}\ and\ \bibinfo
  {author} {\bibnamefont{et~al}},\ }%
  \bibfield{journal}{%
  \bibinfo {journal} {Phys. Rev. Lett.}\ }%
  \textbf{\bibinfo {volume} {107}},\ \bibinfo {pages} {043902} (\bibinfo {year}
  {2011})%
  \bibAnnoteFile{NoStop}{Strelkov}%
\bibitem{LewensteinPhase}%
  \BibitemOpen
  \bibfield{author}{%
  \bibinfo {author} {\bibfnamefont{M.}~\bibnamefont{Lewenstein}}, \bibinfo
  {author} {\bibfnamefont{P.}~\bibnamefont{Sailieres}},\ and\ \bibinfo {author}
  {\bibfnamefont{A.}~\bibnamefont{L'Huillier}},\ }%
  \bibfield{journal}{%
  \bibinfo {journal} {Phys Rev. A}\ }%
  \textbf{\bibinfo {volume} {52}},\ \bibinfo {pages} {4747} (\bibinfo {year}
  {1995})%
  \bibAnnoteFile{NoStop}{LewensteinPhase}%
\bibitem{ShinPhaseIntensity}%
  \BibitemOpen
  \bibfield{author}{%
  \bibinfo {author} {\bibfnamefont{H.~J.}\ \bibnamefont{Shin}}\ and\ \bibinfo
  {author} {\bibnamefont{et~al}},\ }%
  \bibfield{journal}{%
  \bibinfo {journal} {Phys Rev. A}\ }%
  \textbf{\bibinfo {volume} {63}},\ \bibinfo {pages} {1050} (\bibinfo {year}
  {2001})%
  \bibAnnoteFile{NoStop}{ShinPhaseIntensity}%
\end{thebibliography}%

\end{document}